\documentclass[%
 reprint,
 superscriptaddress,
 floatfix,
 amsmath,amssymb,
 aps,prl
]{revtex4-1}
\usepackage{graphicx}
\usepackage{dcolumn}
\usepackage{bm}
\usepackage{hyperref}

\begin{document}

\title{Micrometer-scale ballistic transport of electron pairs in LaAlO\texorpdfstring{$_3$}{3}/SrTiO\texorpdfstring{$_3$}{3} nanowires}

\author{Michelle Tomczyk}
\affiliation{Department of Physics and Astronomy, University of Pittsburgh, Pittsburgh, PA 15260, USA}
\affiliation{Pittsburgh Quantum Institute, Pittsburgh, PA, 15260 USA}
\author{Guanglei Cheng}
\affiliation{Department of Physics and Astronomy, University of Pittsburgh, Pittsburgh, PA 15260, USA}
\affiliation{Pittsburgh Quantum Institute, Pittsburgh, PA, 15260 USA}
\author{Hyungwoo Lee}
\affiliation{Department of Materials Science and Engineering, University of Wisconsin-Madison, Madison, WI 53706, USA}
\author{Shicheng Lu}
\affiliation{Department of Physics and Astronomy, University of Pittsburgh, Pittsburgh, PA 15260, USA}
\affiliation{Pittsburgh Quantum Institute, Pittsburgh, PA, 15260 USA}
\author{Anil Annadi}
\affiliation{Department of Physics and Astronomy, University of Pittsburgh, Pittsburgh, PA 15260, USA}
\affiliation{Pittsburgh Quantum Institute, Pittsburgh, PA, 15260 USA}
\author{Joshua P. Veazey}
\altaffiliation[Present Address: ]{Department of Physics, Grand Valley State University, Allendale, MI 49401}
\affiliation{Department of Physics and Astronomy, University of Pittsburgh, Pittsburgh, PA 15260, USA}
\author{Mengchen Huang}
\affiliation{Department of Physics and Astronomy, University of Pittsburgh, Pittsburgh, PA 15260, USA}
\affiliation{Pittsburgh Quantum Institute, Pittsburgh, PA, 15260 USA}
\author{Patrick Irvin}
\affiliation{Department of Physics and Astronomy, University of Pittsburgh, Pittsburgh, PA 15260, USA}
\affiliation{Pittsburgh Quantum Institute, Pittsburgh, PA, 15260 USA}
\author{Sangwoo Ryu}
\affiliation{Department of Materials Science and Engineering, University of Wisconsin-Madison, Madison, WI 53706, USA}
\author{Chang-Beom Eom}
\affiliation{Department of Materials Science and Engineering, University of Wisconsin-Madison, Madison, WI 53706, USA}
\author{Jeremy Levy}
 \email[Corresponding Author: ]{jlevy@pitt.edu}
\affiliation{Department of Physics and Astronomy, University of Pittsburgh, Pittsburgh, PA 15260, USA}
\affiliation{Pittsburgh Quantum Institute, Pittsburgh, PA, 15260 USA}

\date{\today}

\begin{abstract}
High-mobility complex-oxide heterostructures and nanostructures offer new opportunities for extending the paradigm of quantum transport beyond the realm of traditional III-V or carbon-based materials. Recent quantum transport investigations with LaAlO$_3$/SrTiO$_3$-based quantum dots have revealed the existence of a strongly correlated phase in which electrons form spin-singlet pairs without becoming superconducting. Here we report evidence for micrometer-scale ballistic transport of electron pairs in quasi-one-dimensional (quasi-1D) LaAlO$_3$/SrTiO$_3$ nanowire cavities. In the paired phase, Fabry-Perot-like quantum interference is observed, in sync with conductance oscillations observed in the superconducting regime (at zero magnetic field). Above a critical magnetic field $B_p$, electron pairs unbind and conductance oscillations shift with magnetic field. These experimental observations extend the regime of ballistic electronic transport to strongly correlated phases. 
\end{abstract}

\maketitle

SrTiO$_3$-based heterostructures \cite{Ohtomo2004} and nanostructures \cite{Irvin2013} host a wide range of physical phenomena, including magnetism \cite{Brinkman2007} and superconductivity \cite{Reyren2007}. In particular, LaAlO$_3$/SrTiO$_3$ (LAO/STO) heterostructures exhibit strong, tunable spin-orbit coupling \cite{BenShalom2010,Caviglia2010}, a cascade of structural transitions \cite{Muller1979}, and non-trivial interactions between ferroelastic domain boundaries \cite{Kalisky2013,Honig2013}. LAO/STO-based nanowires possess further surprising behaviors, including \textit{intrinsic} quasi-1D superconductivity \cite{Veazey2013a}, and strong electron pairing outside of the superconducting regime \cite{Cheng2015}. Compared with the 2D superconductor-insulator transition, the nature of correlated electron transport in clean 1D systems remains largely unexplored.

Unlike ballistic semiconducting counterparts, STO-based heterostructures exhibit a relatively short phase coherence, of order $\sim$100\,\,nm \cite{Rakhmilevitch2010,Lin2013}. However, there is growing evidence that within quasi-1D LAO/STO-based channels, scattering lengths, both elastic and inelastic, may be drastically enhanced. Transport measurements of $\sim$10\,\,nm-wide channels at the LAO/STO interface show a nearly two-order-of-magnitude enhancement of Hall mobility, which extends to room temperature \cite{Irvin2013}. Quasi-1D LAO/STO nanowires exhibit conductance values that hover near the single-channel conductance quantum $e^2/h$, independent of channel length \cite{Cheng2013}. There have been stronger claims that the appearance of conductance steps in edge-defined LAO/STO quantum wires implies ballistic transport \cite{Ron2014}. However, conductance steps can arise from any point-like constriction \cite{Wees1988}, and do not imply long-range coherent or ballistic transport. Such conductance steps have also been reported in top-gated STO structures that do not possess a 1D geometry \cite{Gallagher2014}.

Quantum interference experiments can provide useful information about electron scattering. Analogous to photonic interference in an optical Fabry-Perot cavity, multiple reflections of electrons from the endpoints of a nanowire cavity can lead to strong interference effects when the elastic scattering length exceeds the cavity length. This interference requires not only phase coherence but also absence of scattering \cite{Kretinin2010}; many systems with long coherence lengths have much shorter elastic scattering lengths. In ballistic Fabry-Perot cavities, the conductance through the cavity oscillates as a function of the Fermi wavelength, which varies with the chemical potential and is usually controlled by a nearby gate electrode. Only a few material systems have been shown to be capable of supporting micrometer-scale quantum interference: suspended single-wall carbon nanotubes \cite{Liang2001}, high-mobility graphene structures \cite{Miao2007}, and stacking-fault-free III-V nanowires grown by vapor-liquid-solid techniques \cite{Kretinin2010}. 
However, these systems often operate in a regime where electron correlations can be neglected; exceptions include Wigner crystal phases, and magnetically and structurally confined one-dimensional systems (i.e., Tomonaga-Luttinger liquids \cite{Haldane1981}).

In this Letter, we observe evidence of long-range ballistic transport of electron pairs in a complex oxide system. This constitutes a new regime in which strong electronic correlations combine with ballistic electron transport, which is the basis for a remarkable variety of quantum transport phenomena \cite{Nazarov2009}, to achieve greater functionality.

\begin{figure}
\includegraphics[]{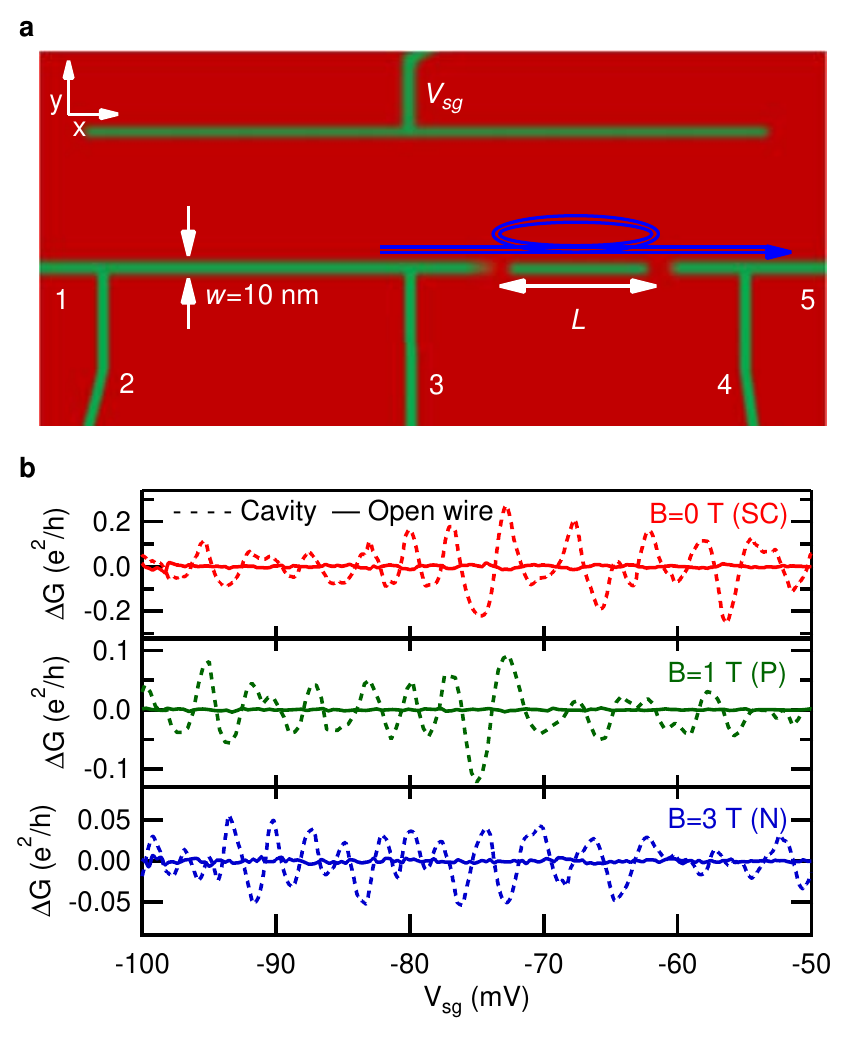}
\caption{\label{fig:fig1}Device schematic and Fabry-Perot oscillations. 
(a) Schematic of cavity device defined by two barriers separated by length $L$. 
Interference due to coherent scattering in the cavity results in conductance 
oscillations periodic in Fermi momentum. 
(b) Background-subtracted zero-bias differential conductance 
($dI/dV$) of the cavity [between voltage leads 3 and 4 in 
(a)] and the open wire (between leads 2 and 3) in the superconducting 
(red), paired (green), and normal (blue) phases of Device A clearly reveals 
large oscillations are only present in the cavity.}
\end{figure}

To investigate the ballistic nature of transport in LAO/STO nanostructures, quasi-1D Fabry-Perot cavities are created at the LAO/STO interface using conductive atomic force microscope (c-AFM) lithography \cite{Cen2008}. A c-AFM tip is placed in contact with and moved across the LAO surface. Positive voltages applied to the tip locally switch the LAO/STO interface to a conductive state (“write”), while negative voltages applied to the tip locally restore the LAO/STO interface to an insulating state (“erase”). To create the geometry shown in Fig.\,\,\ref{fig:fig1}(a), first a nanowire of width $w\approx 10$\,\,nm is written, followed by erasure steps to create semitransparent barriers at both ends of the cavity. Devices are transferred to a dilution refrigerator within 5 minutes of writing to minimize decay, and are cooled to a base temperature $T=50$\,\,mK for transport measurements. Current flows through the main channel containing the two barriers. An applied side gate voltage $V_{sg}$ tunes both the transparency of the barriers and the Fermi level in the cavity. Independent voltage leads enable four-terminal measurements of the cavity conductance, as well as that of an adjoining “open” nanowire, i.e., without barriers. The differential conductance is extracted numerically from $I-V$ curves measured as a function of $V_{sg}$  and magnetic field. Lock-in measurements are performed at reference frequency $f=13.46$\,\,Hz and amplitude $100\,\,\mu$V. Cavities of length $L=0.25-4\,\,\mu$m were studied, and all show qualitatively similar behavior. Additional details of sample growth and fabrication of the nanowire and barriers are described elsewhere (see Supplemental Material \cite{[{}][{, which includes Refs. [25-36], for details on sample growth, nanolithography, modeling of Fabry-Perot interference, background-subtraction calculations, calculations of critical current, and single barrier devices.}]Supp}).


There are three distinct transport regimes \cite{Cheng2015} as a function of the applied magnetic field: superconducting (SC), paired (P), and normal (N). At temperatures below $T_c\approx 300$\,\,mK, and for out-of-plane magnetic fields below $B_c = \mu_0 H_{c2} \approx 0.2$\,\,T, the LAO/STO interface exhibits a sharp increase in conductance that is attributed to superconductivity, both for 2D heterostructures \cite{Reyren2007} and 1D nanowires \cite{Veazey2013a}. The regime $B_c < B < B_p$ has been previously identified as a strongly correlated phase in which electrons exist as spin-singlet pairs without forming a superconducting condensate \cite{Cheng2015}. At sufficiently large magnetic fields (above $B_p \approx 2-5$\,\,T), electrons are unpaired and behave “normally”.

As a function of $V_{sg}$, typical differential conductance $G = dI/dV$ measurements of the cavity exhibit quasi-periodic oscillations at zero-bias, i.e., $V_{4T} = 0$\,\,V. The variation in conductance $G$ after subtraction of a slowly-varying background (see Supplemental Materials \cite{Supp} for details) shows clear oscillations in the cavity, but not in the open wire, in all three phases [Fig.\,\,\ref{fig:fig1}(b)]. In the superconducting state, the conductance oscillations correspond to modulation of the critical current \cite{Supp}.

\begin{figure}
\includegraphics[]{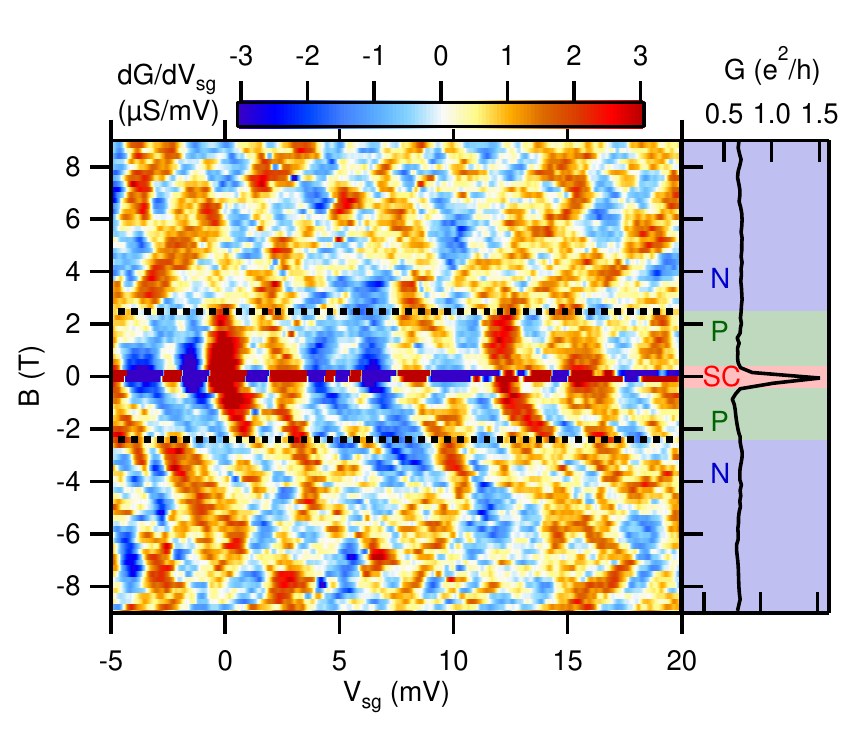}
\caption{\label{fig:fig2}Magnetic field dependence of conductance oscillations. Left, Transconductance $dG/dV_{sg}$ from a lock-in amplifier measurement of $G$ at small (100\,\,$\mu$V) bias versus $B$ and $V_{sg}$ for Device B. Alternating red and blue regions correspond to conductance oscillations. Right, A linecut of $G$ versus $B$ at $V_{sg}=0$\,\,mV shows a sharp peak attributed to superconductivity at $|B| < B_c \approx 0.2$\,\,T (shaded red), while the conductance in the paired (shaded green) and normal (shaded blue) phases is reduced.}
\end{figure}

The transconductance $dG/dV_{sg}$ (Fig.\,\,\ref{fig:fig2}, left), which is computed by numerically differentiating the zero-bias conductance $G$ with respect to side gate, reveals distinct features in the superconducting, paired and normal regimes. The superconducting state is characterized by a sharp conductance peak below $B < B_c$, (Fig.\,\,\ref{fig:fig2}, right, shaded red); correspondingly, the transconductance exhibits large oscillations. For $B > B_c$, the oscillations decrease in amplitude, yet remain in phase with the superconducting state modulation, confirming that transport continues to be dominated by electron pair states despite the loss of superconducting coherence. The phase of the oscillations is preserved over the magnetic field range $B_c < B < B_p$ (shaded green), indicating an overall insensitivity to magnetic fields, consistent with the spin-singlet nature of the paired state. For $B > B_p$ (shaded blue), the electron pairs break and the transconductance oscillations shift markedly with magnetic field. 

\begin{figure}
\includegraphics[]{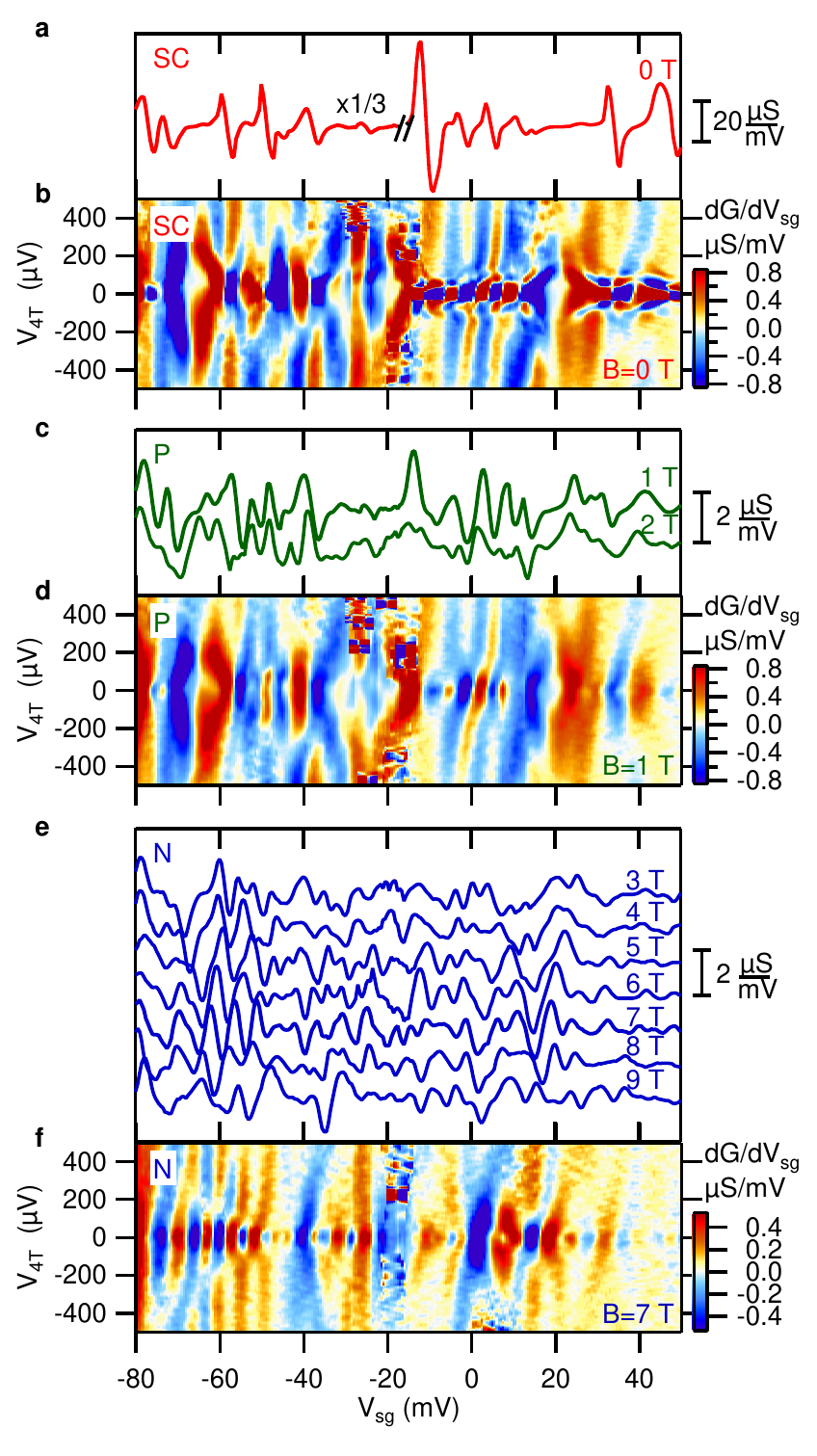}
\caption{\label{fig:fig3}Fabry-Perot interference signatures at finite bias for an $L = 4 \mu$m cavity.  (a), (c), (e), Zero-bias $dG/dV_{sg}$ of Device C as a function of $V_{sg}$ at magnetic fields from 0\,\,T (red) to 9\,\,T (blue) in 1\,\,T steps (each curve offset for clarity). The left half of the red curve at $B = 0$\,\,T has been amplified by a factor of 3 compared to the right half. (b), (d), (f), $dG/dV_{sg}$ vs $V_{4T}$ and $V_{sg}$ in the superconducting phase ((b), $B=0$\,\,T), paired phase ((d), $B=1$\,\,T) and normal, unpaired electron phase ((f), $B=7$\,\,T), corresponding to the respective zero-bias linecuts.}
\end{figure}

The observed transconductance oscillations are consistent with Fabry-Perot interference in cavity devices up to 4\,\,$\mu$m in length (Fig.\,\,\ref{fig:fig3}). Transmission resonances through the cavity occur when the quantum phase associated with round-trip passage is altered by a change in chemical potential or magnetic (Zeeman) interaction. In the “equilibrium” case [Fig.\,\,\ref{fig:fig3}(a),(c),(e)], in which there is no net bias across the cavity, oscillations appear as a function of the applied side gate voltage, which changes the wavelength of the propagating electron states. In the “non-equilibrium” regime [Fig.\,\,\ref{fig:fig3}(b),(d),(f)], an applied source-drain bias can also change the phase; the result is a characteristic “checkerboard” pattern similar to what has been reported for other systems such as carbon nanotubes \cite{Liang2001,Kretinin2010}. Non-equilibrium effects such as heating and intermode scattering can dephase transport and damp the oscillations at sufficiently high source-drain bias values.

The observation of Fabry-Perot interference in the paired regime provides evidence for ballistic transport of electron pairs in the quasi-1D LAO/STO nanowire system. This result is in sharp contrast to Cooper pair insulators, in which electron pairs surviving outside of the superconducting state are localized \cite{Phillips2003}. Metallic Bose phases have been observed in both optical lattice \cite{Deissler2010} and solid state \cite{Phillips2003} systems, but even in clean superconductors where the mean free path is longer than the superconducting coherence length, the mean free path is only on the order of 10 nm \cite{Tsen2015}. Additionally, these metallic Bose phases always appear below the upper critical field for superconductivity in the systems. The results observed here in LAO/STO nanowires are distinct due to both the ballistic nature of transport of the uncondensed electron pairs, and the persistence of this ballistic pair state well above the upper critical field for supconductivity in LAO/STO. 

While conductance oscillations through the cavity are evident for all values of magnetic field explored (up to 9 T), the open wire shows strong suppression of oscillations in all three phases (Fig.\,\,\ref{fig:fig1}\,(b)). The root-mean-square amplitude of conductance fluctuations of the open wire is reduced by an order of magnitude compared with the cavity, suggesting that imperfections in the nanowires contribute negligibly to scattering. The pattern of behavior described here, for both cavities and open wires, is consistently observed for all of the 50 cavity devices studied.

Devices with a single manufactured barrier, in which no interference is expected to occur, were also studied, and typical behavior is shown in Supplementary Fig. S7 \cite{Supp}. Above a conductance value of $\sim e^2/h$, the conductance increases monotonically with increasing gate bias, showing no signs of Fabry-Perot interference.  In some devices, Fabry-Perot signatures are observed; however, in each of those devices, the low-$V_{sg}$ regime also shows quantum dot signatures \cite{Cheng2015}. These signatures are consistent with the existence of a second, unintentional potential barrier along the nanowire that creates a cavity and associated interference patterns. 

While systems which support Fabry-Perot interference are expected to act as quantum dots when tuned to a tunneling regime, not all 1D quantum dot systems can exhibit Fabry-Perot interference \cite{Buitelaar2002}. Resonant tunneling observed in LAO/STO nanowire-based quantum dots at low $V_{sg}$ suggests that extended coherent states exist \cite{Cheng2015}, but does not rule out disorder, which randomizes carrier paths in the transport regime at high $V_{sg}$. In contrast, observation of Fabry-Perot interference as described here demonstrates micrometer-scale elastic scattering lengths in quasi-1D LAO/STO nanowires.

The detailed nature of the observed Fabry-Perot oscillations depends not only on the physical dimensions of the cavity, but on the band structure of the material \cite{Wang2000}. Resonant transmission through a cavity of length $L$ is periodic in the Fermi momentum, $k_F=n\pi/L$, so that the period is inversely proportional to length; however, a quadratic relationship between $k_F$ and Fermi energy $E_F$ leads to a resonance period which depends on the effective mass of the energy band, and increases with energy (see Fig.\,\,S2) \cite{Supp}. Bulk STO has three degenerate 3$d$ conduction bands with $t_{2g}$ orbital character, and interfacial confinement produces an approximately 50\,\,meV upward shift of the $d_{xz}$ and $d_{yz}$ bands relative to the lighter $d_{xy}$ band \cite{Salluzzo2009}. Additionally, the finite width of the quasi-1D nanowire can introduce a manifold of transverse subbands. When new subbands become accessible, abrupt changes in oscillation frequency are expected and observed, further obscuring a direct relationship between device length and the interference $V_{sg}$ period.

Observation of signatures of ballistic transport in quasi-1D LAO/STO nanowires in both the normal- and paired-electron phases contrasts with behavior reported in 2D devices. However, understanding the distinctive transport in quasi-1D structures is possibly relevant for transport measurements of the 2D LAO/STO interface, where local probes have revealed the existence of narrow channel flow along ferroelastic domain boundaries \cite{Kalisky2013,Honig2013}. Additionally, one-dimensional transport offers rich physics with many theoretical \cite{Giamarchi2004} predictions, including charge/spin separation \cite{Haldane1981}. Long-range coherent and ballistic transport in a strongly-correlated electronic phase, along with the reconfigurable nature of this interface system, extend the ability to design novel quantum materials. 

\begin{acknowledgments}
This work was supported by ARO MURI W911NF-08-1-0317 (J.L.), AFOSR MURI FA9550-10-1-0524 (C.B.E., J.L.), FA9550-12-1-0342 (C.B.E.), and FA9550-15-1-0334 (C.B.E.), and grants from the National Science Foundation DMR-1104191 (J.L.), DMR-1124131 (C.B.E., J.L.) and DMR-1234096 (C.B.E.).
\end{acknowledgments}

%
%
%
\end{document}


\title{Supplemental Materials for ``Micrometer-scale ballistic transport of electron pairs in LaAlO\texorpdfstring{$_3$}{3}/SrTiO\texorpdfstring{$_3$}{3} nanowires"}


\author{Michelle Tomczyk}
\affiliation{Department of Physics and Astronomy, University of Pittsburgh, Pittsburgh, PA 15260, USA}
\affiliation{Pittsburgh Quantum Institute, Pittsburgh, PA, 15260 USA}
\author{Guanglei Cheng}
\affiliation{Department of Physics and Astronomy, University of Pittsburgh, Pittsburgh, PA 15260, USA}
\affiliation{Pittsburgh Quantum Institute, Pittsburgh, PA, 15260 USA}
\author{Hyungwoo Lee}
\affiliation{Department of Materials Science and Engineering, University of Wisconsin-Madison, Madison, WI 53706, USA}
\author{Shicheng Lu}
\affiliation{Department of Physics and Astronomy, University of Pittsburgh, Pittsburgh, PA 15260, USA}
\affiliation{Pittsburgh Quantum Institute, Pittsburgh, PA, 15260 USA}
\author{Anil Annadi}
\affiliation{Department of Physics and Astronomy, University of Pittsburgh, Pittsburgh, PA 15260, USA}
\affiliation{Pittsburgh Quantum Institute, Pittsburgh, PA, 15260 USA}
\author{Joshua P. Veazey}
\altaffiliation[Present Address: ]{Department of Physics, Grand Valley State University, Allendale, MI 49401}
\affiliation{Department of Physics and Astronomy, University of Pittsburgh, Pittsburgh, PA 15260, USA}
\author{Mengchen Huang}
\affiliation{Department of Physics and Astronomy, University of Pittsburgh, Pittsburgh, PA 15260, USA}
\affiliation{Pittsburgh Quantum Institute, Pittsburgh, PA, 15260 USA}
\author{Patrick Irvin}
\affiliation{Department of Physics and Astronomy, University of Pittsburgh, Pittsburgh, PA 15260, USA}
\affiliation{Pittsburgh Quantum Institute, Pittsburgh, PA, 15260 USA}
\author{Sangwoo Ryu}
\affiliation{Department of Materials Science and Engineering, University of Wisconsin-Madison, Madison, WI 53706, USA}
\author{Chang-Beom Eom}
\affiliation{Department of Materials Science and Engineering, University of Wisconsin-Madison, Madison, WI 53706, USA}
\author{Jeremy Levy}
 \email[Corresponding Author: ]{jlevy@pitt.edu}
\affiliation{Department of Physics and Astronomy, University of Pittsburgh, Pittsburgh, PA 15260, USA}
\affiliation{Pittsburgh Quantum Institute, Pittsburgh, PA, 15260 USA}

\date{\today}

\maketitle

\renewcommand{\theequation}{S\arabic{equation}}
\renewcommand{\thefigure}{S\arabic{figure}}
\renewcommand{\thetable}{S\arabic{table}}
\renewcommand{\bibnumfmt}[1]{[S#1]}
\renewcommand{\citenumfont}[1]{S#1}

\section{\label{sec:growth}Sample growth and preparation}
LaAlO$_3$/SrTiO$_3$ (LAO/STO) samples are grown by pulsed laser deposition (PLD) \cite{Park2010S,Bark2011S,Bark2012S}. The STO substrate is TiO$_2$ terminated by etching in buffered HF for 60 seconds, and annealed at 1000$^{\circ}$C for 6 hours to achieve an atomically smooth surface. A thin (3.4 unit cell) LAO film is subsequently grown on top of STO by PLD at a temperature of 550$^{\circ}$C and 1x10$^{-3}$\,\,mbar oxygen pressure, and gradually cooled to room temperature. Electrical contact to the LAO/STO interface is made by Ar$^+$ etching (25\,\,nm) followed by sputter deposition of Ti/Au (5 nm/20 nm). Additional details are described in Ref.\,\,\cite{Cheng2011S,Levy2014S}.

\section{\label{sec:lithography}C-AFM nanolithography and barrier creation}
For LAO/STO samples grown with an LAO thickness just below the critical thickness of 4\,\,u.c. at which the two-dimensional conducting layer appears, the interface is insulating and easily tunable by either back gate or top gate through the metal-insulator transition (MIT) \cite{Thiel2006S}. A conductive AFM (c-AFM) tip can be used as a top gate to locally induce the MIT at the interface by applying a positive voltage to the surface. Moving the positively-biased ($V\sim+10$ V) tip across the surface creates conducting structures less than 10\,\,nm wide \cite{Cen2008S} at the interface. These nanowires can be ‘erased’ and the interface returned to an insulating state by moving a negatively biased ($V\sim-10$ V) c-AFM tip across the surface of the LAO. When a very small negative voltage (-0.05\,\,V\,\,$<V_{tip}<$\,\,-0.5\,\,V) is applied to the tip, and the tip is moved perpendicularly across an existing nanowire, a nanoscale potential barrier is created in the wire. The size of the barrier is characterized by monitoring the change in resistance during the cutting process at room temperature. The cavity lengths $L$ between the barriers ranged from 250\,\,nm to 4\,\,$\mu$m. The distance from each barrier to the nearest voltage lead was held constant for all devices at 750\,\,nm. The total distance between voltage leads (leads 3 and 4, Fig.\,\,1a in the main text) was therefore $L+1.5\,\,\mu$m. The 4-terminal voltage (leads 2 and 3, Fig.\,\,1a in the main text) of a segment of nanowire equal in length to the total $L+1.5\,\,\mu$m, but without manufactured barriers, was also measured as a control, as parameters such as $V_{sg}$ and $B$ were varied. The side gate was created with the same c-AFM lithography as the device, running parallel to the main current-carrying channel, about 1\,\,$\mu$m away. Additional details are described in Ref.\,\,\cite{Cheng2015S}.

\section{\label{sec:data}Representative \texorpdfstring{$dI/dV$}{dI/dV} of six devices}
\begin{figure}
\includegraphics[]{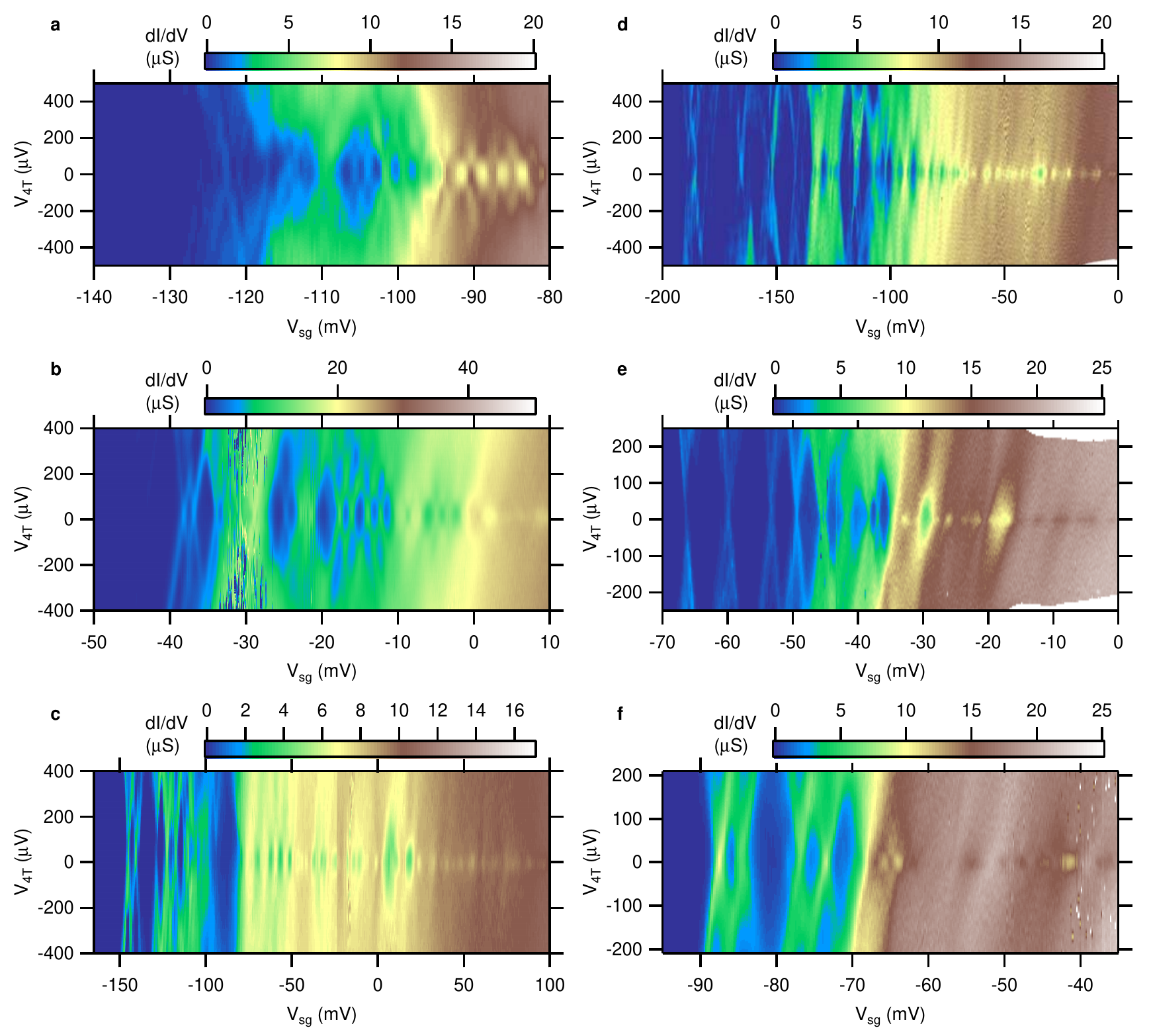}
\caption{\label{fig:figS1}Fabry Perot interference in many devices. a-f, Extended $dI/dV$ of the cavity for Devices A-F, respectively, at $T$=50 mK. At low $V_{sg}$, the cavity is in the blockade regime and diamonds are present. At higher $V_{sg}$, the devices become conducting and Fabry-Perot oscillations are observed.}
\end{figure}

The side gate tunes both the Fermi level in the cavity and the transparency of the barriers. At very negative side gates, the nanowire cavity is very weakly coupled to the leads, resulting in diamond-shaped regions of conductance blockade (Fig.\,\,\ref{fig:figS1}). This is indicative that the device is behaving as a quantum dot, with resonant tunneling only when the chemical potential of the leads aligns with an energy level on the dot. At more positive gate voltages, the barriers become more transparent. When the Fermi level in the cavity rises above the potential barriers, the device becomes conducting and the barriers act as the primary scattering centers. Long coherence and scattering lengths in the nanowires enable Fabry-Perot interference effects. Sometimes a range of $V_{sg}$ between the blockaded transport and interference oscillations exhibits a crossover regime, in which oscillations are occasionally interrupted by weak blockade, or vice versa. The full, normal-state tunability of the differential conductance ($dI/dV$) of Devices A, B and C in the main text, along with three other devices, is seen in Fig.\,\,\ref{fig:figS1}. Interference oscillations in $dI/dV$ appear in the conducting state of all devices, with cavity lengths ranging up to 4 microns. The device parameters are given in Table\,\,\ref{tab:tabS1}.

\begin{table}
\begin{ruledtabular}
\begin{tabular}{llll}
Device Name & Cavity Length $L$ ($\mu$m) & Back gate (V) & Magnetic field in Fig. \ref{fig:figS1} (T) \\
A & 0.25 & -1 & 3.0\\
B & 1.0 & 0 & 4.0\\
C & 4.0 & -3 & 7.0\\
D & 1.0 & -2 & 7.0\\
E & 1.0 & 0 & 1.0\\
F & 0.5 & -0.7 & 6.0\\
\end{tabular}
\end{ruledtabular}
\caption{\label{tab:tabS1}Device Parameters. Summary of cavity length $L$ between the manufactured barriers, backgate voltage applied during measurements, and the applied magnetic field for the data in Fig. \ref{fig:figS1}, for Devices A-F.}
\end{table}

Fabry-Perot interference is expected to appear as periodic dips in a high conductance regime \cite{Liang2001S}, where the cavity is strongly coupled to the leads. The preeminence of dips, rather than peaks, has been explained by inter-mode coupling at the scattering centers \cite{Liang2001S}. The participation of multiple subbands within the cavity increases the likelihood of inter-mode scattering, which can lead to suppression of coherence signatures. At finite bias, a range of available momenta could also suppress coherence signatures. This is likely why full ‘checkerboard’ patterns only appear in small subsets of gate voltage in most devices, despite all cavity devices exhibiting zero-bias conductance oscillations.

\section{\label{sec:FPmodel}Modeling of Fabry-Perot Interference}
\begin{figure}
\includegraphics[]{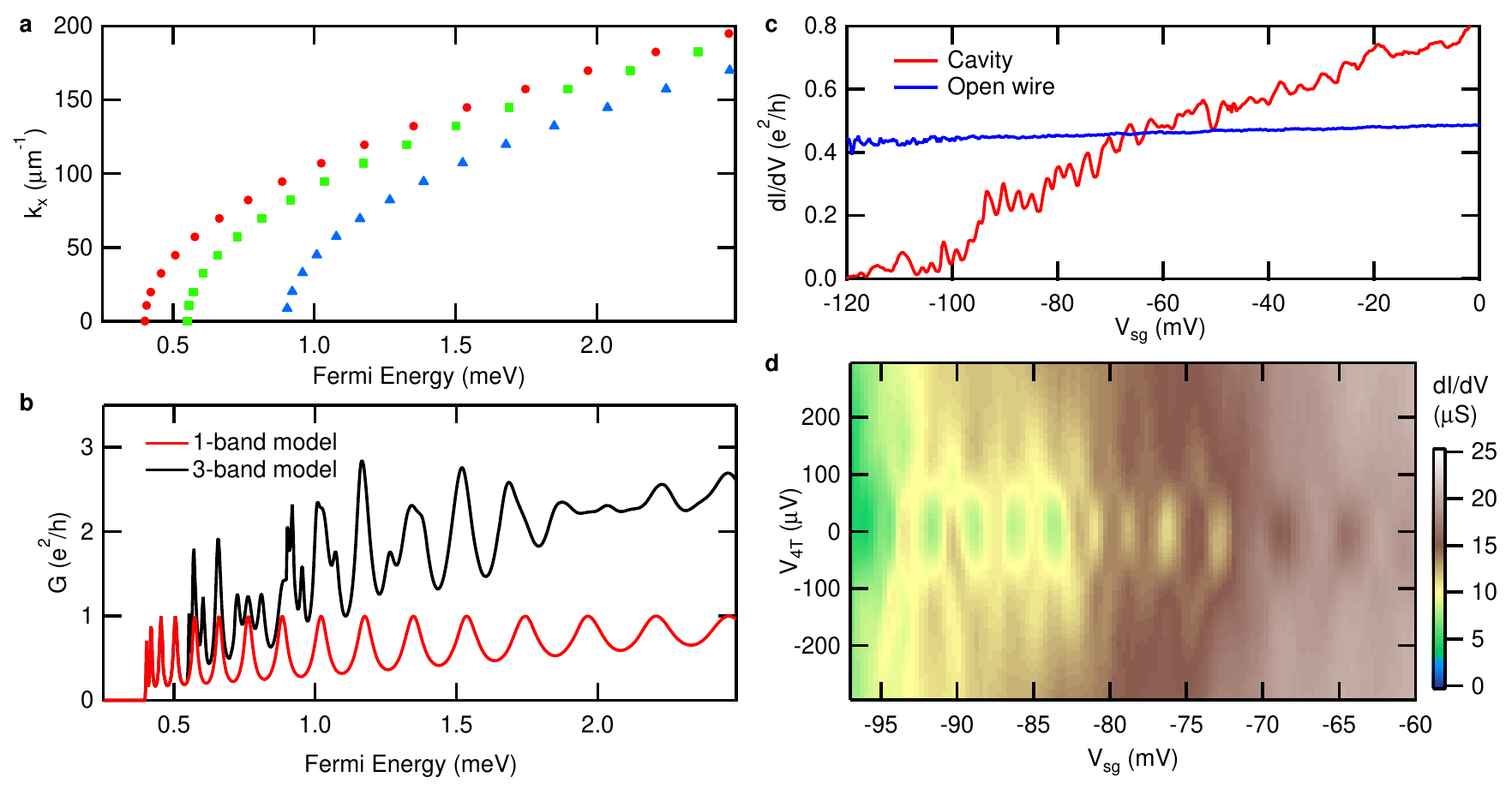}
\caption{\label{fig:figS2}Semi-classical transmission model. 
a, Resonant states periodic in momentum are depicted by symbols for parabolic dispersion of three bands. b, Conductance oscillations due to the lowest energy band in (a) (red) and conductance oscillations due to coherent transport of all three bands depicted in (a) (black). c, Zero-bias ($V_{4T}=0$) differential conductance ($dI/dV$) of Device A ($L=0.25\,\,\mu$m, $B=3$\,\,T) for both the cavity and the open wire. Cavity conductance features quasi-periodic oscillations that qualitatively resemble a multimode transmission model. d, $dI/dV$ versus $V_{4T}$ and $V_{sg}$ for Device A show a smoothly-changing period over a subset of $V_{sg}$.}
\end{figure}

Both the geometry of the device and the band structure of the material contribute to the interference signatures in a Fabry-Perot cavity \cite{Wang2000S,Connor1968S}.  For materials with a single band, resonant transmission through a cavity of length $L$ is periodic in the Fermi momentum, $k_F=n\pi/L$. While a linear  dependence of Fermi energy $E_F$ on momentum leads to a constant $V_{sg}$ resonance period, a quadratic energy dispersion leads to a $V_{sg}$ period which depends on the effective masses of the various bands, and increases with energy \cite{Wang2000S} (“1-band model” in Fig.\,\,\ref{fig:figS2}\,a,b). Bulk STO has three degenerate 3$d$ conduction bands with $t_2g$ orbital character. Interfacial confinement produces a $\sim$50\,\,meV upward shift of the $d_{xz}/d_{yz}$ bands relative to the $d_{xy}$ band \cite{Salluzzo2009S}, while lateral confinement in quasi-1D nanowires is expected to create a manifold of transverse subbands.  Fig.\,\,\ref{fig:figS2}\,b shows an expected interference pattern for a nanowire with three distinct subbands.

Total conductance is calculated from the Landauer formula

\begin{equation}\label{eq:cond}
  G=\frac{e^2}{h}\sum_i T_i
\end{equation}

where $T_i$ are the transmission of each energy band $i$. In this analysis, each band is assumed to contribute $e^2/h$, not 2$e^2/h$, because the simulation is being compared with data taken in large magnetic fields which drive the LAO/STO interface system normal and break electron pairs \cite{Cheng2015S}, so that energy bands are not assumed to be spin degenerate. Transmission in a quasi-classical approximation \cite{Connor1968S} is given by 

\begin{equation}\label{eq:trans}
\begin{split}
  T_i=\frac{1}{P^2+Q^2+PQcos2k_iL};\\
  P(\epsilon_L,\epsilon_R)=\sqrt[]{(1+e^{-2\pi\epsilon_L})(1+e^{-2\pi\epsilon_R})};\\
  Q(\epsilon_L,\epsilon_R)=e^{-\pi(\epsilon_L+\epsilon_R)};\\
\end{split}
\end{equation}

where $\epsilon_{L,R}=(E_F-V_b)/\hbar\omega$ give the Fermi energy $E_F$ normalized by the barrier height $V_b$ and width $\omega$. At each $E_F$, the momentum $k_i$ for each band $i$ with band bottom $E_i$ below $E_F$ was calculated for a parabolic dispersion 

\begin{equation}\label{eq:ki}
  k_i=\frac{\sqrt{2m_{eff}(E_F-E_i)}}{\hbar}
\end{equation}

The three-band model in Fig.\,\,\ref{fig:figS2} uses an effective mass \cite{Santander2011S} $m_{eff}=0.7m_e$ for all bands, $E_1=300\,\,\mu$eV, $E_2=550\,\,\mu$eV and $E_3=900\,\,\mu$eV, barrier height $V_b=100\,\,\mu$eV, barrier width $\omega=1\text{x}10^{13}\,\,s^{-1}$, and length of the cavity $L=250$\,\,nm. The momentum states which give a maximum in $T_i$ are shown in Fig.\,\,\ref{fig:figS2}\,a for the energy dispersion in Eq.\,\,(\ref{eq:ki}). For the lowest band depicted (red circles), the conductance in units of $e^2/h$ is calculated according to Eqs.\,\,(\ref{eq:cond}-\ref{eq:trans}). Since Fig.\,\,\ref{fig:figS2}\,a-b share an axis, it is easy to see that each resonant state in the dispersion of the lowest (red) band in Fig.\,\,\ref{fig:figS2}\,a corresponds to a peak in conductance in the red curve in Fig.\,\,\ref{fig:figS2}\,b. The resonant states occur periodically in $k_i$, and therefore the spacing between resonances increases as a function of $E_F$. Finally, the conductance for all three bands was calculated according to Eqs.\,\,(\ref{eq:cond}-\ref{eq:trans}) (Fig.\,\,\ref{fig:figS2}\,b, black). In this case, beating between the resonances occurs, resulting in what appear to be random fluctuations in conductance. Zero-bias $dI/dV$ linecuts in the normal, unpaired state (Fig.\,\,\ref{fig:figS2}\,c) clearly show the qualitative similarity between the multiband model and the conductance oscillations observed in cavity devices, contrasted with the lack of such features in the “open” wires with no barriers. A plot of $dI/dV$ extended to finite bias shows a slowly-increasing period between resonances, as expected, for a small range of $V_{sg}$ (Fig.\,\,\ref{fig:figS2}\,d).

\section{\label{sec:backsub}Background Subtraction}

\begin{figure}[h]
\includegraphics[]{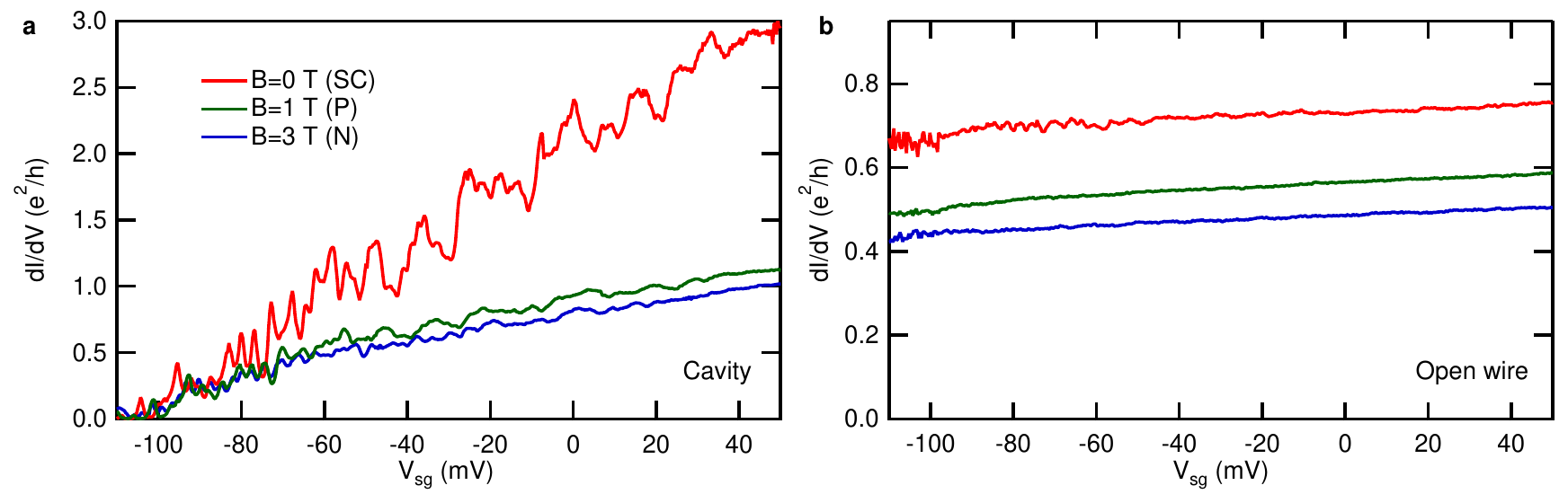}
\caption{\label{fig:figS3}Differential conductance. a-b, Zero-bias differential conductance ($dI/dV$) of the cavity (between voltage leads 3 and 4 in Fig.\,\,1a in the main text) and the open wire (between leads 2 and 3) in the superconducting (red), paired (green) and normal (blue) phases of Device A.}
\end{figure}

\begin{figure}[h]
\includegraphics[]{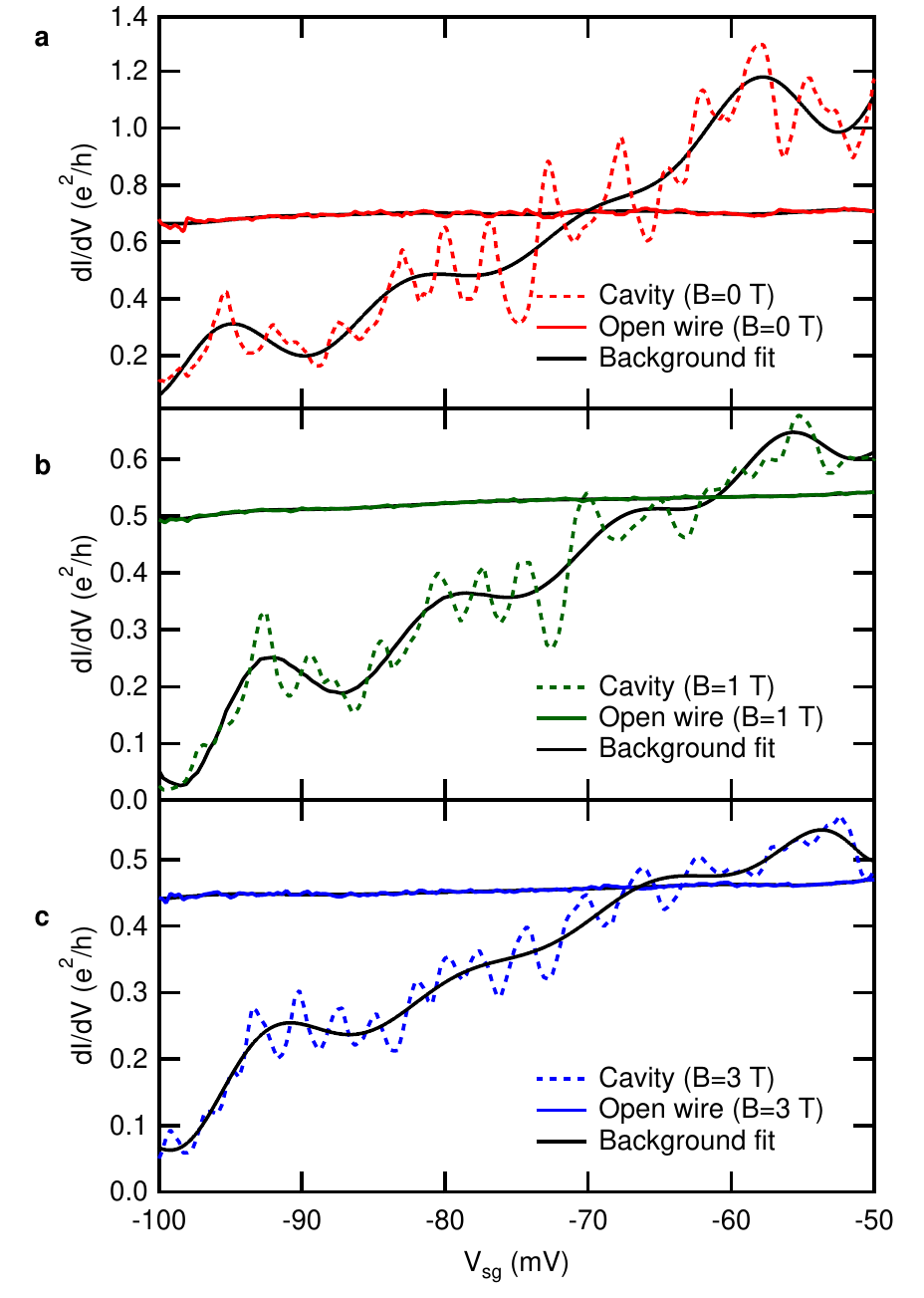}
\caption{\label{fig:figS4}Background subtraction. a-c, $dI/dV$ of the cavity (dash) and open wire (solid color) for Device A at $B=0$\,\,T (a), $B=1$\,\,T (b) and $B=3$\,\,T (c). Data shown here is the subset -100\,\,mV$< V_{sg} <-50$\,\,mV of the corresponding red, green and blue curves in Fig.\,\,\ref{fig:figS3}. A slowly-varying background is overlaid on each curve (solid black). The result $\Delta G$ (Fig.\,\,1(b) in the main text) of subtracting the slowly-varying background from $dI/dV$ reveals Fabry-Perot interference in the cavity.}
\end{figure}

The original zero-bias $dI/dV$ linecuts from which the panels in Fig.\,\,1(b) in the main text were derived are shown in Fig.\,\,\ref{fig:figS3} for both the cavity (a) and the open wire (b). A high-order polynomial fit to a $V_{sg}$ subset from -105 to -48\,\,mV was performed and the resulting slowly-varying background is overlaid with the original data in Fig.\,\,\ref{fig:figS4} for both the cavity and open wire in the superconducting (a), paired (b) and normal (c) phases. The root-mean-square amplitude of the fluctuations in the open wire are suppressed by over 90\% compared to the cavity. Interestingly, the background conductance of the normal-state cavity reveals step-like features superimposed beneath the oscillations, reminiscent of interference oscillations originally predicted in ballistic devices with quantized conductance \cite{Kirczenow1989S}. While these steps are clearly much less than $e^2/h$, this suggests that perhaps, with refinement of these experiments, quantized conductance is possible to achieve in these ballistic LAO/STO nanowires.

\section{\label{sec:critcurrent}Superconducting Phase and Calculation of Critical Current}
\begin{figure}[h]
\includegraphics[]{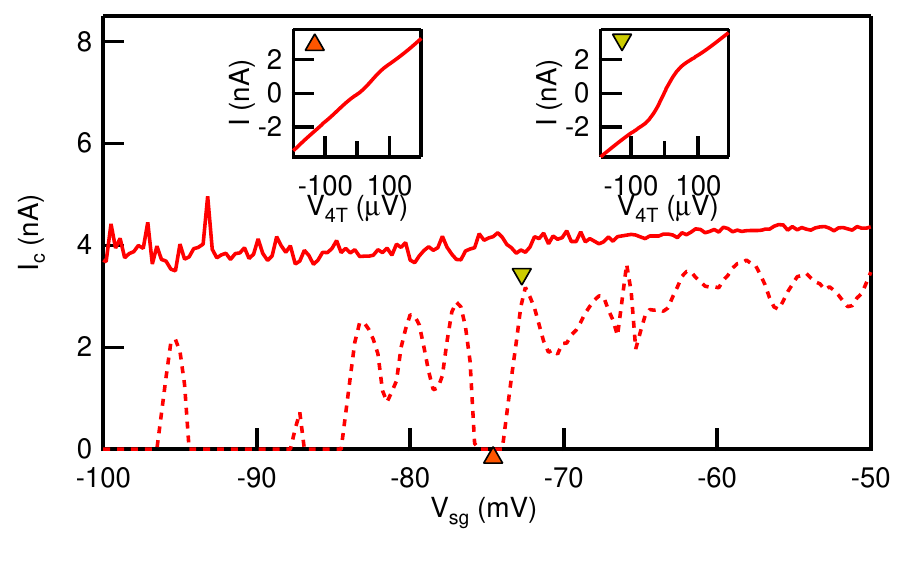}
\caption{\label{fig:figS5}Critical Current Modulations. In the superconducting state, critical current $I_c$ of the cavity can be greatly modulated with $V_{sg}$, while $I_c$ of the open wire is mostly constant. Insets show $I-V$ data resulting in low (triangle) and high (inverted triangle) $I_c$, which also correspond to low and high conductance oscillations.}
\end{figure}

In the regime $|B| < B_c$, the device is superconducting and the conductance is significantly enhanced (Fig.\,\,\ref{fig:figS3}, red) compared to the non-superconducting paired phase (green) and the normal phase (blue). While a zero-resistance superconducting state is usually not achieved in nanowires (insets of Fig.\,\,\ref{fig:figS5}), likely due to the increased susceptibility of low-dimensional superconductors to thermally-activated phase slips and other effects \cite{Veazey2013S}, the nanowire cavity shows a strong enhancement of conductance oscillations in the superconducting regime (Fig\,\,1b). These features are associated with a modulation of the critical current $I_c$ (Fig.\,\,\ref{fig:figS5}), similar to supercurrent transistors \cite{Jarillo2006S}. While such strong $I_c$ modulation does not occur in the open wire, a slight anti-correlation is observed between the $I_c$ of the cavity and open wires (e.g.  80\,\,mV $< V_{sg} <  70$\,\,mV).  

\begin{figure}[h]
\includegraphics[]{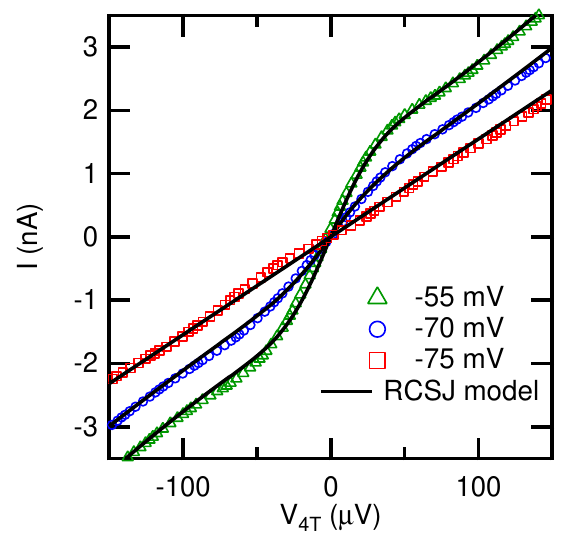}
\caption{\label{fig:figS6}Fitting data to RCSJ model. Typical current $I$ versus voltage $V_{4T}$ curves for the cavity in Device A at $B=0$\,\,T, at three $V_{sg}$ values in the range shown in Fig.\,\,\ref{fig:figS5}. Black lines are fits calculated through the RCSJ model in Eq.\,\,(\ref{eq:critcurr}).}
\end{figure}

To calculate the critical current for the cavity data, a resistive and capacitively shunted junction (RCSJ) model was used based on Ref.\,\cite{Jorgensen2007S}, which models a quantum dot between two superconducting leads. The discrete energy levels due to confinement in the dot also give rise to resonant transmission in the Fabry-Perot regime. Starting with an overdamped RCSJ model with a sinusoidal current-phase relation, Ref.\,\cite{Jorgensen2007S} includes current due to Andreev reflections and finds a current-voltage ($I-V_{sd}$) relation

\begin{equation}\label{eq:critcurr}
\begin{split}
  I(V_{sd})=I_cIm\bigg[ \frac{I_{1-i\eta(V_{sd})}(I_c\hbar/2ek_BT)}{I_{-i\eta(V_{sd})}(I_c\hbar/2ek_BT)}\bigg]+\frac{V_J(V_{sd})}{R_J};\\
  V_J(V_{sd})=V_{sd}-RI(V_{sd});\\
  \eta(V_{sd})=\hbar V_{sd}/2eRk_BT;\\
\end{split}
\end{equation}

where $I_\alpha(x)$ is the modified Bessel function. The lead resistance $R$, the resistance carrying the Andreev current $R_J$, and the critical current $I_c$ are the free parameters, for $T=50$\,mK. $R_J$ is assumed to be larger than $R$, and this is satisfied by the fits. Examples of fits to the data which result in $I_c$ versus $V_{sg}$ for the cavity in Fig.\,\,\ref{fig:figS5} are shown in Fig.\,\,\ref{fig:figS6}. 

The RCSJ model is not appropriate for the open wire, however, because there should be no Andreev reflections, and therefore we expect $R_J<R$. Indeed, fitting the open wire $I-V_{sd}$ curves to Eq.\,\,(\ref{eq:critcurr}) results in $R_J<R$, invalidating the fit. Instead, we use a simpler definition. $I_c$ for the open wire is defined as the location of the resistance peaks in the $dV/dI$ versus $I$ curve.

\section{\label{sec:singlebarrier}Single-barrier devices}
\begin{figure}[h]
\includegraphics[]{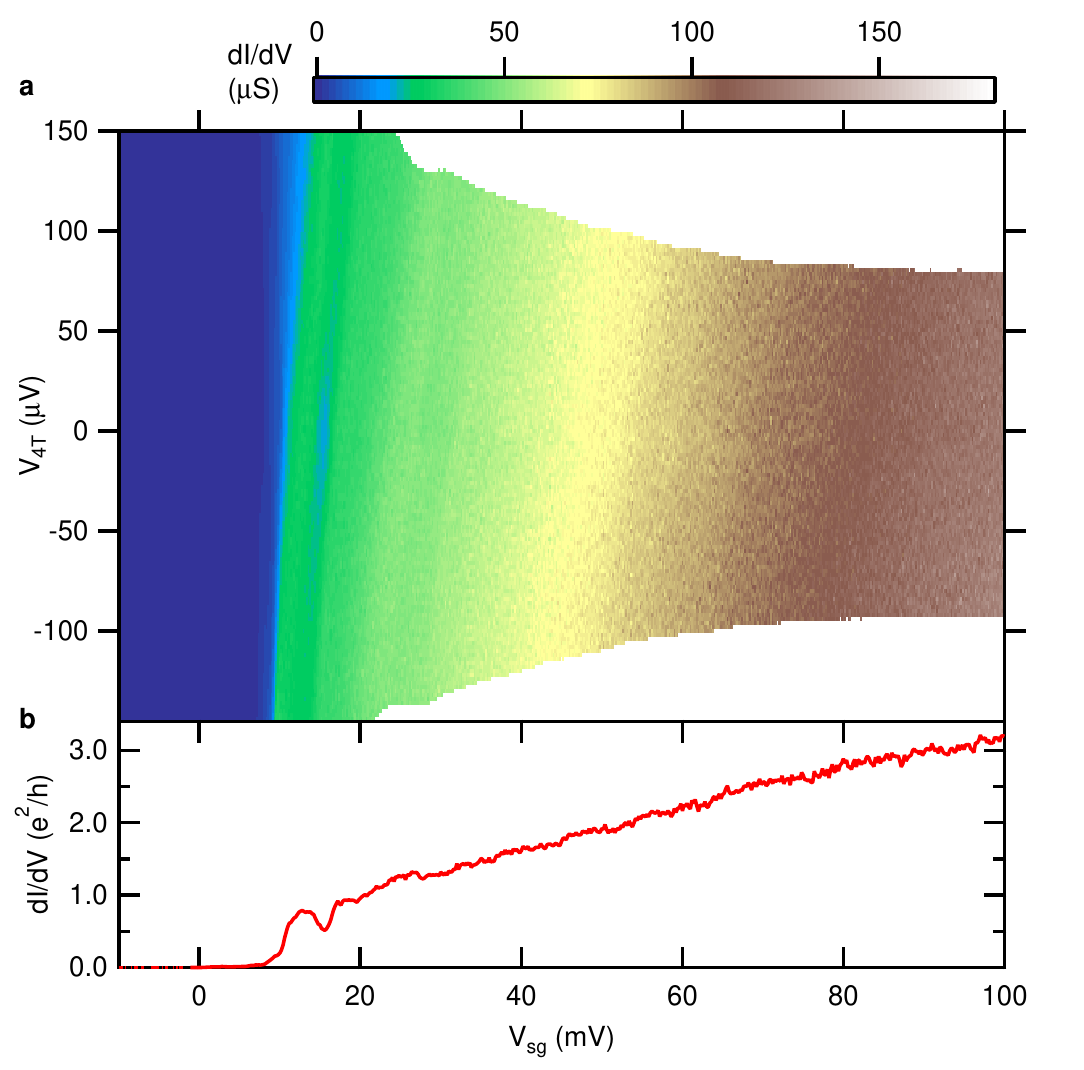}
\caption{\label{fig:figS7}Single barrier device. a, $dI/dV$ versus $V_{4T}$ and $V_{sg}$ for Device G at $B=3$\,\,T. b, $dI/dV$ linecut at zero-bias ($V_{4T}=0$). No Fabry Perot conductance oscillations are observed.}
\end{figure}

Twelve devices were made with a single barrier, rather than the two barriers which define a cavity. The four-terminal leads were between 0.5-1.5\,\,$\mu$m from the barrier, for a total wire length of 1-3\,\,$\mu$m between the leads for the dozen devices. Half of the devices show no blockade or Fabry-Perot, like Device G in Fig.\,\,\ref{fig:figS7}. Compared to a cavity device, such as Fig.\,\,\ref{fig:figS2}\,c, Device G clearly has no quasi-periodic oscillations like those observed in Devices A-F (Fig.\,\,\ref{fig:figS1}), even at zero-bias. The only non-linear behavior occurs as the device is pinched off by a low side gate. The other half of the single barrier devices exhibit both blockade behavior and Fabry-Perot interference, suggesting that an unintentional second barrier exists, forming a cavity. These potential barriers may contribute additional features in some of the devices with two engineered barriers. However, the disorder is not strong enough to cause blockade or Fabry-Perot signatures in open nanowires with no intentionally manufactured barriers, further supporting the claim of a long elastic scattering length.

